\documentstyle[aaspp4,12pt]{article}

\newcommand{\etal}{{\em et~al. }}
\newcommand{\Msun}{{M$_\odot$ }}

\newcommand{\vsini}{{$v\sin i$ }}

\begin{document}

\title{The Structural Effects of Rotation in Low Mass Stars}
\author{Alison Sills, M. H. Pinsonneault}
\affil{Department of Astronomy, The Ohio State University, 174 W. 18th
Ave., Columbus, OH, 43210, USA}

\begin{abstract}
We present theoretical models of rotating low mass stars (0.1 - 1.0
\Msun) to demonstrate the effect of rotation on the effective
temperature and luminosity of stars. The range of rotation rates in
our models corresponds to the observed rotation rates in young low
mass stars. Rotation lowers the effective temperature and luminosity
of the models relative to standard models of the same mass and
composition. We find that the decrease in T$_{eff}$ and L can be
significant at the higher end of our mass range, but becomes small
below 0.4 \Msun. The effects of different assumptions about internal
angular momentum transport are discussed. Formulae for relating
T$_{eff}$ to mass and v$_{rot}$ are presented. We demonstrate that the
kinetic energy of rotation is not a significant contribution to the
luminosity of low mass stars.
\end{abstract}


\section{Introduction}

We are now able to observe stars with regularity down to the hydrogen
burning limit in open clusters (e.g. NGC 2420, von Hippel \etal 1996),
globular clusters (e.g. 47 Tucanae, Santiago, Elson \& Gilmore 1996),
and the field (e.g. Tinney, Mould \& Reid 1993). We have also been
able to observe rotation in these stars, using spectroscopy (Kraft
1965, Stauffer \etal 1997) to determine \vsini, and also using
photometry to monitor spot modulation on the stars (Barnes \& Sofia
1999, Prosser \etal 1995) and thereby determine rotational
periods. This plethora of information about the rotation of low mass
stars has been a great boon to the study of these stars for a number
of reasons. First of all, the rotation rates of stars on the main
sequence are determined by their pre-main sequence evolution, so that
by studying the rotational evolution of low mass stars, we can
investigate the early stages of stellar evolution. Secondly, stellar
rotation is tied to stellar magnetic phenomena. The evolution of the
rotation rates is largely determined by angular momentum loss from a
magnetized stellar wind (Kawaler 1988, Weber \& Davis 1967). Stellar
rotation is found to correlate with chromospheric activity and other
magnetic tracers (for a review see Hartmann \& Noyes 1987), which
lends support to the idea that rotation plays a crucial role in the
generation of stellar magnetic fields, perhaps through a dynamo.

The major obstacles which prevented us from modeling very low mass
stars accurately in the past have been the lack of adequate model
atmospheres, opacities and equations of state for low temperatures
(less than 4000 K). Lately, however, several groups (Alexander \&
Ferguson 1994; Allard \& Hauschildt 1995; Saumon, Chabrier \& Van Horn
1995) have made breakthroughs in the necessary physics. Improved
evolutionary models of very low mass stars have been produced over the
last few years (Baraffe \etal 1998). However, the effects of rotation
have not been included in these recent low mass models, and neglecting
it could lead to anomalous results. For example, rotation can modify
the amount of lithium depletion in low mass stars, affecting the
derived ages from lithium isochrone fitting (e.g. Stauffer, Schultz \&
Kirkpatrick 1998). In this work, we present the first rotational
models of stars less massive than 0.5 \Msun and examine the structural
effects of rotation in these stars. In section 2, we present the
method used to determine the rotational evolution of the low mass
stars. We present the results in section 3, and discuss their
implications in section 4.

\section{Method}

We used the Yale Rotating Stellar Evolution Code (YREC) to construct
models of the low mass stars. YREC is a Henyey code which solves the
equations of stellar structure in one dimension (Guenther \etal
1992). The star is treated as a set of nested, rotationally deformed
shells. Nuclear reaction rates are taken from Gruzinov \& Bahcall
(1998). The initial chemical mixture is the solar mixture of Grevesse
\& Noels (1993), and our models have a metallicity of Z=0.0188.  We
use the latest OPAL opacities (Iglesias \& Rogers 1996) for the
interior of the star down to temperatures of $\log T (K) = 4$. For
lower temperatures, we use the molecular opacities of Alexander \&
Ferguson (1994).  For regions of the star which are hotter than $\log
T (K) \geq 6$, we used the OPAL equation of state (Rogers, Swenson \&
Iglesias 1996). For regions where $\log T (K) \leq 5.5$, we used the
equation of state from Saumon, Chabrier \& Van Horn (1995), which
calculates particle densities for hydrogen and helium including
partial dissociation and ionization by both pressure and temperature.
In the transition region between these two temperatures, both
formulations are weighted with a ramp function and averaged.  The
equation of state includes both radiation pressure and electron
degeneracy pressure. For the surface boundary condition, we used the
stellar atmosphere models of Allard \& Hauschildt (1995), which
include molecular and grain effects and are therefore relevant for low
mass stars.  We used the standard B\"{o}hm-Vitense convective mixing
length theory (Cox 1968; B\"{o}hm-Vitense 1958) with $\alpha$=1.72.
This value of $\alpha$, as well as the solar helium abundance,
$Y_{\odot}=0.273$, was obtained by calibrating models against the
observed radius ($6.9598 \times 10^10$ cm) and luminosity ($3.8515
\times 10^33$ erg/s) at the present age of the Sun (4.57 Gyr).

The structural effects of rotation are treated using the scheme
derived by Kippenhahn \& Thomas (1970) and modified by Endal \& Sofia
(1976). The details of this particular implementation are discussed in
Pinsonneault \etal (1989). In summary, quantities are evaluated on
equipotential surfaces rather than the spherical surfaces usually used
in stellar models. The mass continuity equation is not altered by
rotation:
\begin{equation}
\frac{\partial M}{\partial r} = 4\pi r^2 \rho.
\end{equation}
The equation of hydrostatic equilibrium includes a term which takes
into account the modified gravitational potential of the non-spherical
equipotential surface:
\begin{equation}
\frac{\partial P}{\partial M} = - \frac{GM}{4 \pi r^4} f_P,
\end{equation}
where
\begin{equation}
f_P = \frac{4\pi r^4}{GMS} \frac{1}{\langle g^{-1}\rangle},
\end{equation}
and
\begin{equation}
\langle g^{-1} \rangle = \frac{1}{S} \int_{\psi = const} g^{-1} d\sigma,
\end{equation}
$S$ is the surface area of an equipotential surface, and $d\sigma$ is
an element of that surface. The factor $f_P$ is less than one for
non-zero rotation, and approaches one as the rotation rate goes to
zero.
The radiative temperature gradient also depends on rotation:
\begin{equation}
\frac {\partial \ln T}{\partial \ln P} = \frac{3 \kappa}{16 \pi acG}\frac{P}{T^4}\frac{L}{M}\frac{f_T}{f_P},
\end{equation}
where
\begin{equation}
f_T = \left(\frac{4\pi r^2}{S}\right)^2 \frac{1}{\langle g \rangle \langle g^{-1},
\rangle}
\end{equation}
and $\langle g \rangle$ is analogous to $\langle g^{-1}\rangle$. $f_T$
has the same asymptotic behaviour as $f_P$, but is typically much
closer to 1.0. The energy conservation equation retains its
non-rotating form. Therefore, all the structural effects of rotation
are limited to the equation of hydrostatic equilibrium and the
radiative temperature gradient. This modified temperature gradient is
used in the Schwarzschild criterion for convection:
\begin{equation}
\frac{\partial \ln T}{\partial \ln P} = min \left[\nabla_{ad},\nabla_{rad} \frac{f_T}{f_P} \right]
\end{equation}
where $\nabla_{ad}$ and $\nabla_{rad}$ are the normal spherical adiabatic
and radiative temperature gradients

We start our models on the birthline of Palla \& Stahler (1991), which
is the deuterium-burning main sequence and corresponds to the upper
envelope of T Tauri observations in the HR diagram. It has been shown
(Barnes \& Sofia 1996) that this physically realistic assumption for
the initial conditions of stellar rotation models is crucial for
accurate modeling of ultra-fast rotators in young clusters.  All our
models started with an initial period of 8 days, which corresponds to
the median classical T Tauri star rotation period (Choi \& Herbst
1996).

In this paper, we present stellar models for solar metallicity stars
between 0.1 and 1.0 \Msun in increments of 0.1 \Msun. These models
have been evolved from the birthline to an age of 10 Gyr. We have not
considered any angular momentum loss, since in this paper we wish to
explore the largest reasonable structural changes caused by
rotation. In our standard models, local conservation of angular
momentum is enforced in radiative regions of the star, and convective
zones rotate as solid bodies. We have also calculated models in which
solid body rotation is enforced throughout the star, regardless of the
convective state of the region.

\section{Results}

The initial rotation period of 8 days for our models was chosen
because those models represent the maximum observed rotation rates for
stars in young open clusters. Observed rotation rates as a function of
effective temperature are plotted in figure 1, along with our rotating
models at the zero-age main sequence. The data show here are from five
young clusters: IC 2602 and IC 2391 (30 Myr), $\alpha$ Persei (50
Myr), Pleiades (70 Myr) and Hyades (600 Myr), and were obtained from
the Open Cluster Database (Prosser \& Stauffer). To transform between
the observed V-I colours and T$_{eff}$, we used the empirical colour
calibration of Yuan \etal (1999).  Our models are rotating at or
slightly above the fastest rotation rates observed in these
representative clusters.

The evolutionary tracks for both rotating and non-rotating models are
presented in figure 2. As expected (Sackmann 1970, Pinsonneault \etal
1989), the effect of rotation is to shift stars to lower effective
temperatures and lower luminosities, mimicking a star of lower
mass. This effect is most pronounced for the highest mass stars
presented in this paper, and is reduced to a low level for stars less
than 0.4 \Msun. Since low mass stars are fully convective, their
temperature gradient will be the adiabatic gradient, which does not
depend on rotation rate (equation 7). However, the structural effects
of rotation are still apparently in fully convective stars, and
diminish with decreasing mass. This suggests that an additional
mechanism is also at work.  As stars get less massive, their central
pressure is being provided less by thermal pressure and more by
degeneracy pressure. The amount of degeneracy is determined by the
density in the interior, which is not affected by rotation (see
equation 1). Rotation provides an additional method of support for the
star, but in stars with a significant amount of degeneracy, the
rotational support is a smaller fraction of the total
pressure. Therefore, the structure of the low mass stars is less
affected by rotation than their higher mass counterparts.

Figure 3 compares the evolutionary tracks for rotating stars under
different assumptions about internal angular momentum transport. The
solid tracks are stars which have differentially rotating radiative
cores and rigidly rotating convection zones, while the dashed lines
show the tracks for stars which are constrained to rotate as solid
bodies. The two tracks for each mass have the same surface rotation
rate at the zero-age main sequence. The low mass stars show no
difference between the two assumptions, since these stars are fully
convective for the entire 10 Gyr plotted here. Therefore, they always
rotate as solid bodies. The higher mass stars begin their lives high
on the pre-main sequence as fully convective, solid body rotators. As
they contract and develop radiative cores, however, the difference in
the two assumptions about angular momentum transport becomes
apparent. Differential rotators have a higher total angular momentum
than solid body rotators of the same surface rotation rate. As stars
contract along the pre-main sequence, they become more centrally
concentrated, which means that the core spins up more than the
envelope does.  The solid body rotators are forced to spread their
angular momentum evenly throughout the star, so they have less total
angular momentum for a given surface rotation rate.  Therefore, the
impact of rotation on the structure of the star is larger for
differential rotators than for solid body rotators of the same surface
rotation rate.  However, at constant initial angular momentum, the
solid body rotators are cooler at the zero age main sequence, and have
longer pre-main sequence lifetimes, than differentially rotating stars
of the same mass.  When comparing the effects of rotation between
different models, it is important to note whether the comparison is
between stars with the same current surface rotation rate, or with the
same initial angular momentum.

We included the kinetic energy of rotation ($T=\frac{1}{2}I\omega^2$)
in our determination of the total luminosity in each shell of the
star. As the star changes its moment of inertia $I$ and its rotation
rate $\omega$, the resulting change in its rotational kinetic energy
can be included in the energy budget of the star.  Most
implementations of stellar rotation into stellar structure and
evolution neglect this energy since it is expected that the amount of
kinetic energy available is not enough to significantly affect the
evolution of the star. Since very low mass stars have much lower
luminosities than solar-mass stars, but their moments of inertia are
not as significantly lower, it is plausible that the kinetic energy of
rotation would contribute a significant fraction of the total
luminosity of the star.  As shown in figure 4, however, the change in
the kinetic energy of rotation contributes no more than 6\% of the
total luminosity of the star in the 1.0 \Msun model, and that
contribution lasts less than 50 Myr. As expected, the lowest mass star
has the most significant contribution, lasting for about 1 Gyr, but at
4\% or less. The positions of stars in the HR diagram are minimally
affected by the inclusion of this source of energy. The kinetic energy
of rotation reduces the luminosity at any given time by less than 0.02
dex in $\log(L/L_{\odot})$, and usually less than 0.005 dex. The
timescales for evolution are also equally unaffected. The models
presented in this paper, those with no angular momentum loss, will
have the maximum possible effect of rotational kinetic energy. Since
these models show no significant effect, we conclude that the change
in the kinetic energy of rotation is at most a perturbation on the
structure.

The main structural effect of rotation is a reduction in the effective
temperature of stars. Using our tracks, we have quantified the
relationship between rotational velocity and the difference in
effective temperature at the zero age main sequence. In figure 5 we
present this relationship for stars of different masses, and for both
the differentially rotating (solid lines) and solid body models
(dashed lines). For low mass stars, the difference in temperature
caused by rotation is of order a few tens of degrees (and reduces to
less than 10 degrees for stars of 0.2 \Msun). This difference is
therefore negligible. However, the reduction in effective temperature
is larger for the more massive stars, and can reach significant levels
of a few hundred degrees for stars more massive than about 0.6
\Msun. Therefore, when determining masses from observed temperatures
or colours, it is important know how fast these stars are
rotating. The relationship between rotation rate and difference in
effective temperature, for a given stellar mass, is well-fit by a
polynomial. The coefficients for this polynomial at different masses
and under different assumptions of internal angular momentum transport
are given in table 2. It should be noted that while solid body
rotators of the same initial period rotate faster at the zero age main
sequence than differentially rotating stars, the structural effects of
rotation are slightly more pronounced in the differential rotators at
constant rotation speed. Therefore, for a constant rotational
velocity, stars which rotate differentially have a higher angular
momentum than solid body rotators.

For stars of the same mass, rotation reduces the luminosity of stars
as well as their temperatures. The different in luminosity is not as
important as the difference in temperature cause by rapid rotation, as
shown in figure 6. Even for the most extreme case, the difference in
luminosity for a 1.0 \Msun star rotating at 250 km/s is less than 0.12
dex in $\log(L_{\odot})$. While differences of this size will result
in a thicker main sequence of a cluster, it should not affect any
scientific results significantly. Most stars in clusters do not rotate
very fast, so the upper main sequence will be well-defined for any
isochrone fitting or distance determination. Luminosity is used as an
indicator of mass for low mass stars, but since the difference in
luminosity between rapid rotators and non-rotators is very small for
low mass stars, this calibration should not be affected by rotation.

The total effect of rotation is such that the locus of the zero age
main sequence becomes brighter as stars rotate more quickly. The
combination of a significant decrease in temperature with a small
decrease in luminosity for stars of the same mass moves the locus
above the non-rotating main sequence. At a surface rotation rate of
100 km/s, the rotating main sequence is brighter by about 0.01
magnitudes. At 200 km/s, the sequence is brighter by 0.03 magnitudes.
Therefore, we expect to see rapid rotators in clusters lying above the
cluster main sequences by a few hundredths of a magnitude.

Since they are fainter, rapid rotators have slightly longer lifetimes
compared to non-rotating stars of the same mass. The amount of
increase depends on the mass of the star and the rotation rate, but in
the most extreme case (1.0 \Msun rotating at 250 km/s), the difference
in pre-main sequence lifetime is 7\%. For rotation rates less than 100
km/s, the increase in lifetime is less than 1\% for all masses.

\section{Summary and Discussion}

We have investigated the effect of rotation on the structure of low
mass stars. We discussed a number of implications, based on models
which demonstrate the maximum extent of the differences between
rotating and non-rotating models. The most important effect is the
reduction of effective temperature for stars of a given mass. Rapid
rotators are cooler than slow rotators, and so, for stars more massive
than 0.5 \Msun, any relationship between temperature and mass should
take into account the rotation rate of the star. We have shown that
the structural effects of rotation in very low mass stars (less than
$\sim$ 0.5 \Msun) are minimal and can be neglected when interpreting
temperatures and luminosities of these stars from observations.  Table
2 gives the polynomial correction between the effective temperature of
a rotating star and that of a non-rotating star, as a function of
rotation rate and stellar mass.

We have shown that the kinetic energy of rotation is not a significant
contribution to the total luminosity of stars between 0.1 and 1.0
\Msun, and does not change the timescale for evolution on the pre-main
sequence.  This rotational contribution to the total energy of the
star can therefore be neglected in future evolutionary calculations.

Stellar activity can also influence the color-temperature
relationship; because of the well-known correlation between increased
rotation, increased stellar spot coverage, and increased chromospheric
activity this will tend to change the observed position of rapidly
rotating stars in the HR diagram relative to slow rotators.  Different
color indices are affected to different degrees; for example, Fekel,
Moffett, \& Henry (1986) found a systematic departure between the B-V
and V-I colors of active stars.  Stars with more modest activity
levels have a more normal color-color relationship (e.g.  Rucinski
1987).  Active stars tend to be bluer in B-V than in V-I relative to
less active stars.  Fekel, Moffett, \& Henry (1986) treated this as an
infrared excess, but in open clusters such as the Pleiades and Alpha
Per rapid rotators are on or above the main sequence in V-I, but can
be below the zero-age main sequence in B-V (Pinsonneault et
al. 1998). Given the theoretical trends presented here, this suggests
that V-I is a good tracer of temperature, and that B-V is the colour
which is most affected by activity. The difference between effective
temperatures based on B-V and those based on V-I can reach 200 K in
Pleiades stars (Krishnamurthi, Pinsonneault, King, and Sills 1999).

In this work, we have neglected angular momentum loss, since we wish
to demonstrate the maximum effects of rotation on the structure of low
mass stars. However, angular momentum loss is required to explain the
evolution of rotation rates observed by comparing stars of the same
masses in open clusters of different ages (Barnes \& Sofia 1996,
Krishnamurthi \etal 1997).  We have also chosen only two form of
internal angular momentum transport in these models. Our models either
enforced the local conservation of angular momentum in radiative
regions, and enforce solid body rotation in convection zones, or we
assume the entire star rotates as a solid body. We have not
investigated the case in which local conservation of angular momentum
is enforced throughout the star, nor have we included any transport of
angular momentum caused by any internal instabilities. More detailed
models, which include both angular momentum loss and a more
complicated treatment of internal angular momentum transport, are
needed to accurately model the structure and evolution of rotating
stars. However, these complicated models will follow the same trends
described in this work, with effects of the same magnitude or
less. This paper therefore provides a useful guide to the effects of
rotation on the structure of low mass stars.

\acknowledgements This work was supported by NASA grant
NAG5-7150. A. S. wishes to recognize support from the Natural Sciences
and Engineering Research Council of Canada. We would like to
acknowledge use of the Open Cluster Database, as provided by
C.F. Prosser (deceased) and J.R. Stauffer, and which currently may be
accessed at http://cfa-www.harvard.edu/$\sim$stauffer/, or by anonymous ftp
to cfa0.harvard.edu (131.142.10.30), cd /pub/stauffer/clusters/.

\clearpage

\figcaption[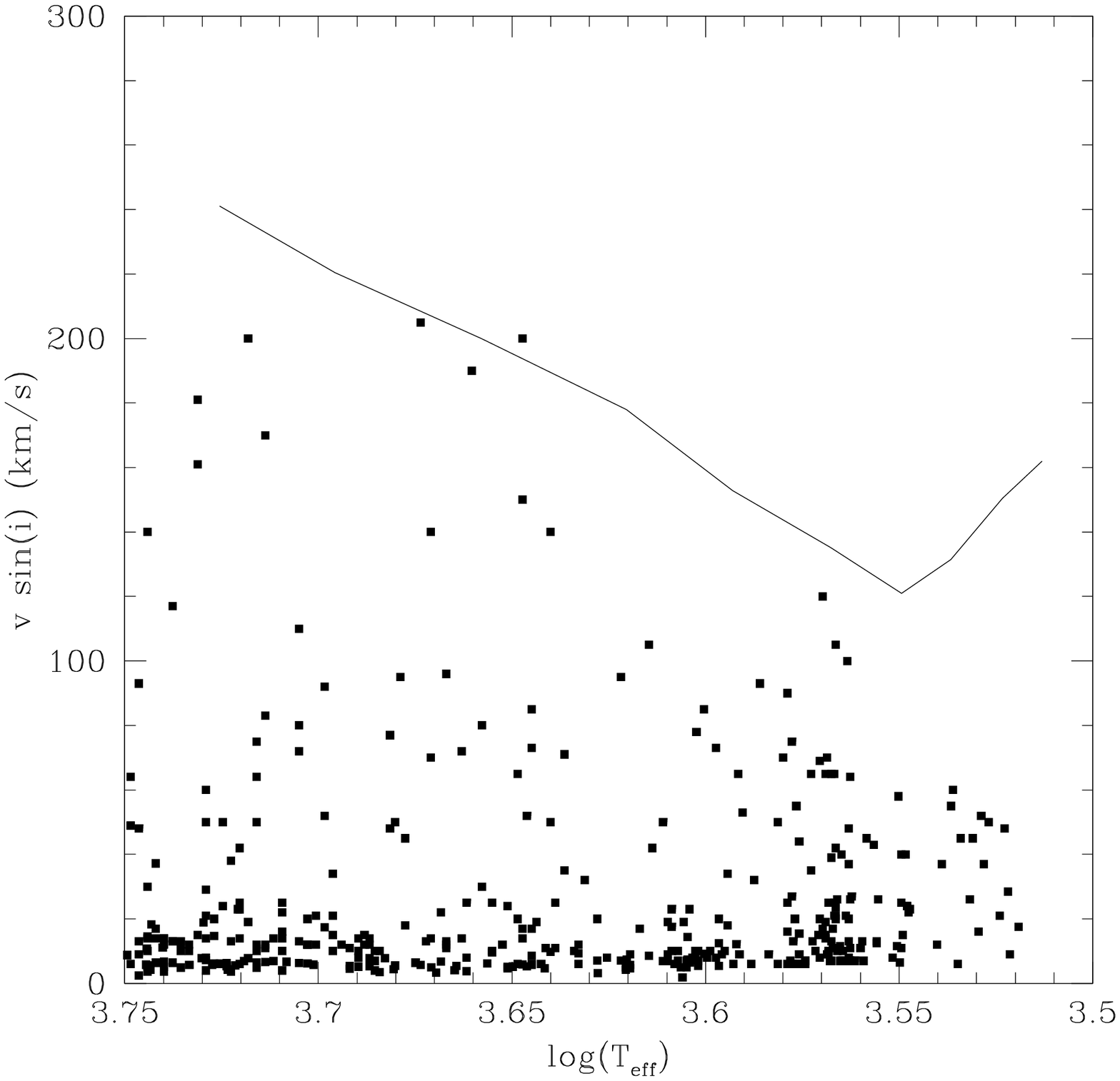]{Observed rotation rates as a function of
temperature for five young open clusters: IC 2602 and IC 2391 (30
Myr), $\alpha$ Persei (50 Myr), Pleiades (70 Myr) and Hyades (600
Myr). The data are from the Open Cluster Database (Prosser \&
Stauffer). The solid line shows differentally rotating models with
initial rotation periods of 8 days and masses between 1.0 and 0.1
\Msun. These models represent the fastest observed rotators in young
open clusters.}

\figcaption[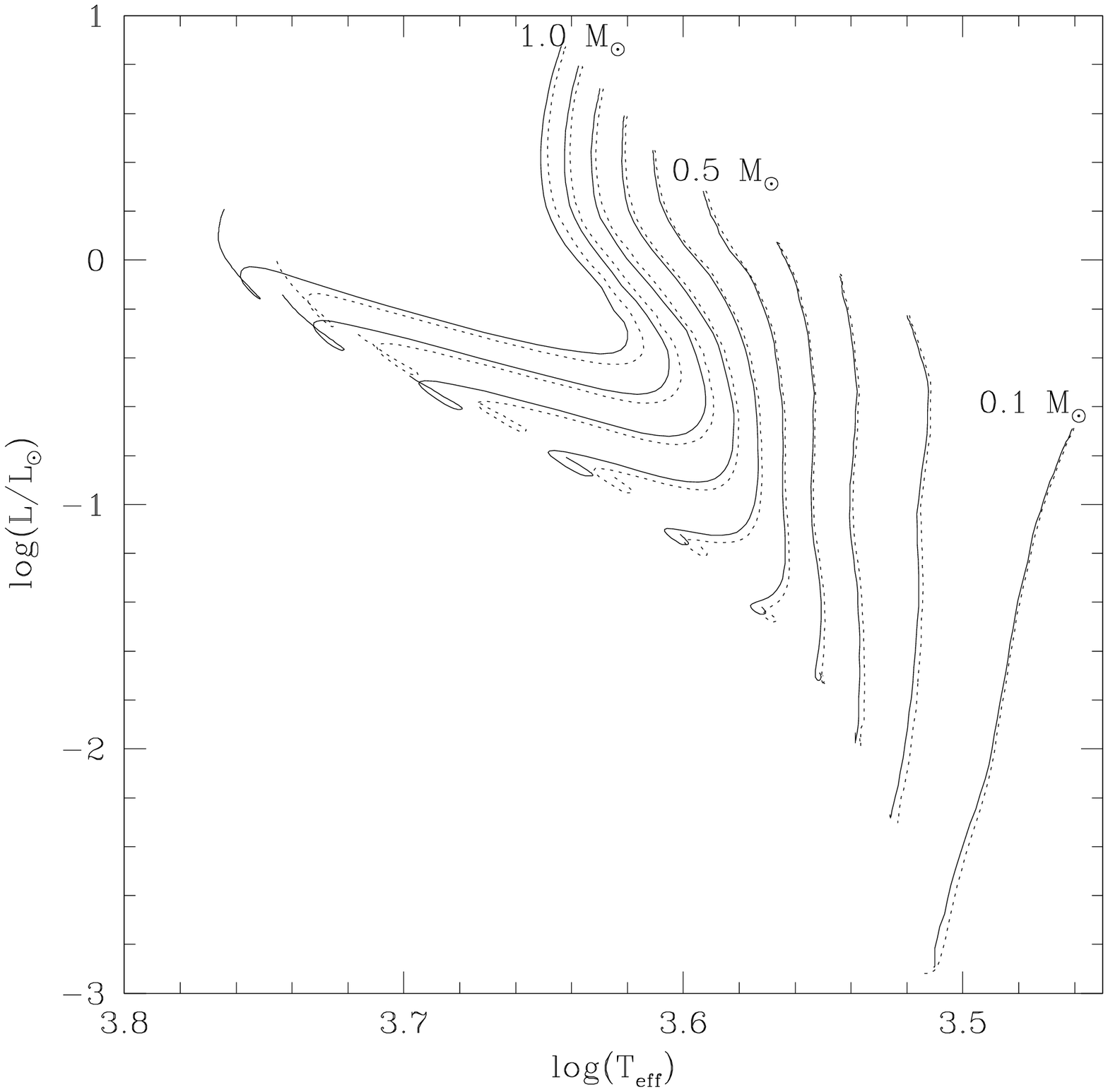]{Evolutionary tracks for stars with masses
between 0.1 \Msun and 1.0 \Msun in steps of 0.1 \Msun. The solid lines
are stars without rotation, and the dotted lines are stars which have
initial rotation periods of 8 days, and no angular momentum loss.}

\figcaption[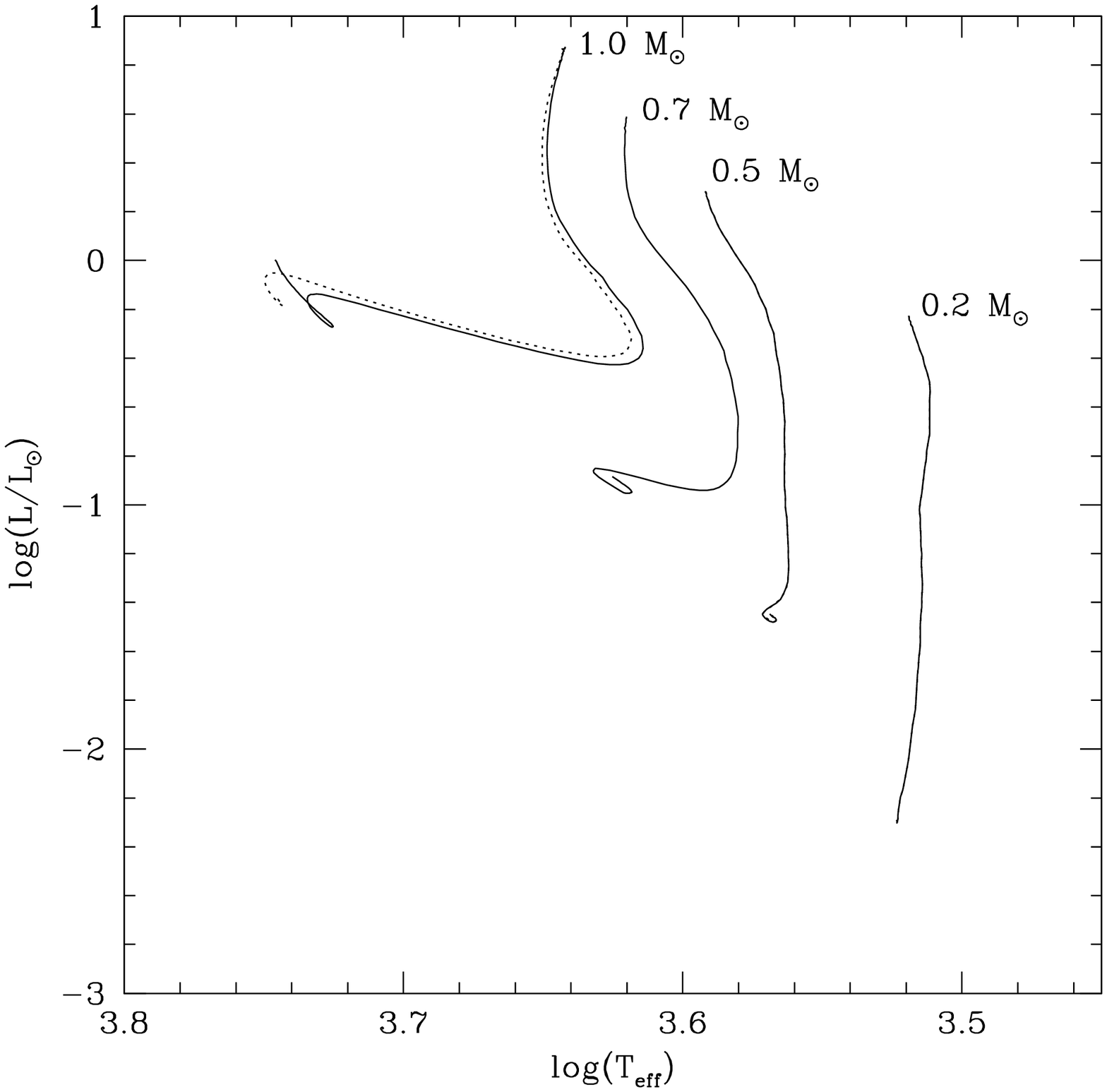]{Evolutionary tracks for rotating stars
under different assumptions about internal angular momentum
transport. The solid tracks are stars which have differentially
rotating radiative cores and rigidly rotating convection zones, while
the dashed lines show the tracks for stars which are constrained to
rotate as solid bodies. Stars of the same mass have the same surface
rotation rate at the zero age main sequence.  Since the low mass stars
are fully convective throughout their pre-main sequence lifetime, they
always rotate as solid bodies, and so the two tracks are
identical. For the higher mass stars, the tracks are the same while
the star is on the early pre-main sequence. When the star begins to
develop a significant radiative core, however, the two cases have
different evolutionary paths.}

\figcaption[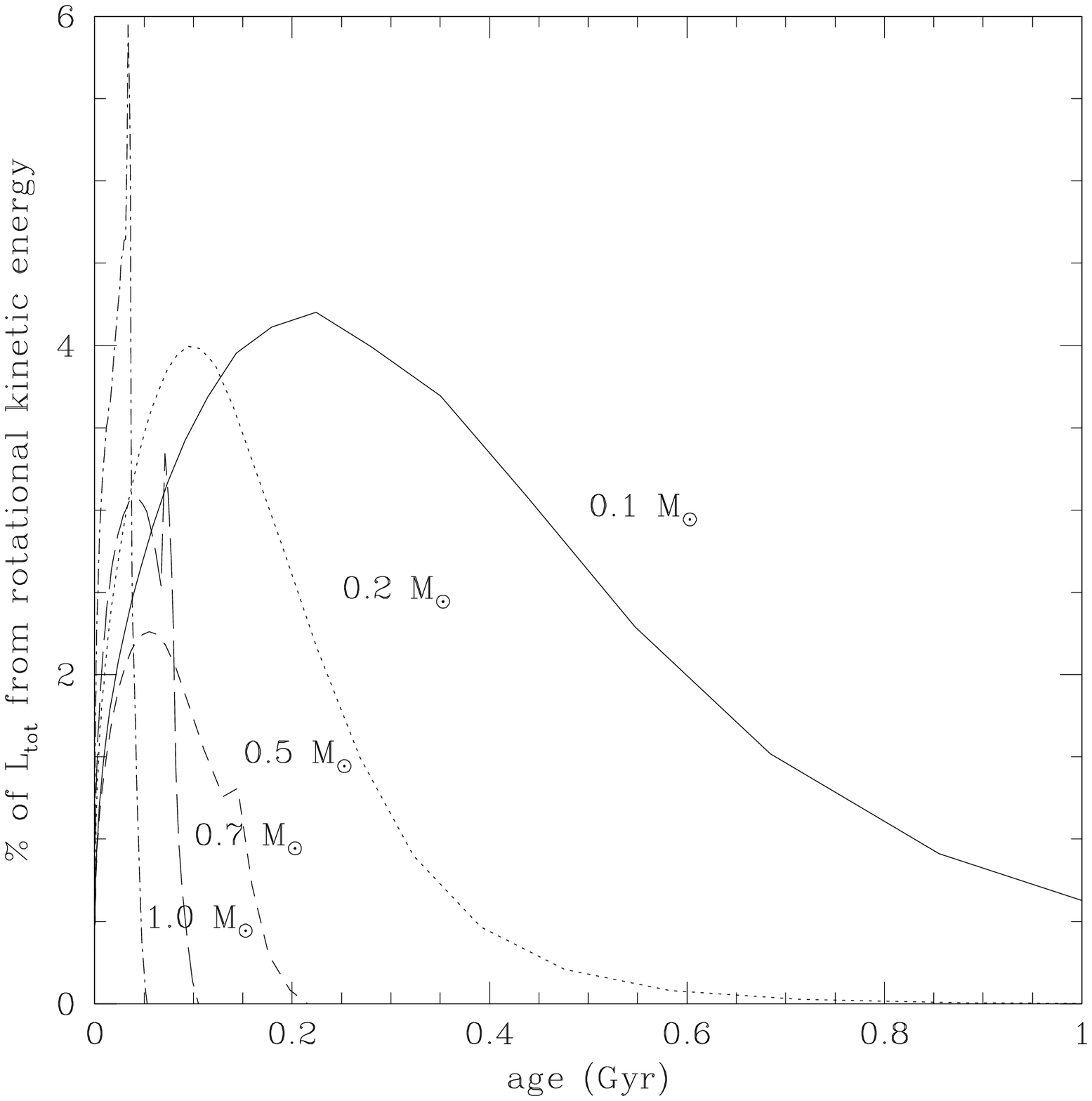]{The percentage contribution to the total
luminosity by the rotational kinetic energy, as a function of age. The
maximum contribution is 6\%, for the 1.0 \Msun star, but for a short
time. For the 0.1 \Msun star, the contribution of rotation kinetic
energy lasts the longest time. In either case, the effect on the
evolutionary timescale and position in the HR diagram is
insignificant.}

\figcaption[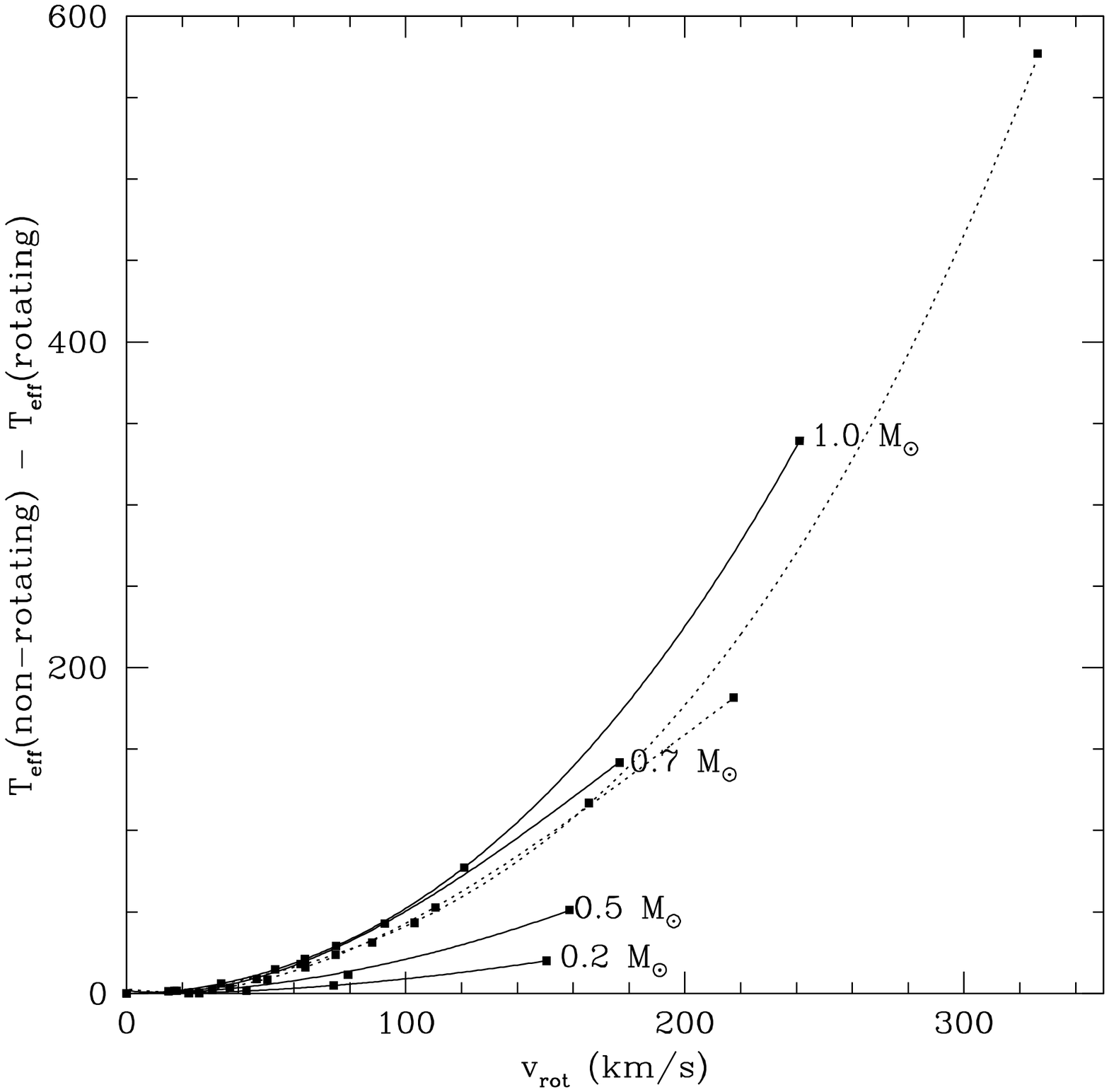]{Difference between effective temperatures of
rotating and non-rotating stars at the same zero age main sequence, as
a function of surface rotation velocity. The different lines
correspond to stars of different masses, with the highest masses
demonstrating the largest difference in temperature. Models which
allow for differential rotation are plotted using solid lines, while
the solid body rotators are plotted as dashed lines.  A difference of
a 100 K or more will significantly affect the mapping of stellar mass
on observed effective temperature.}

\figcaption[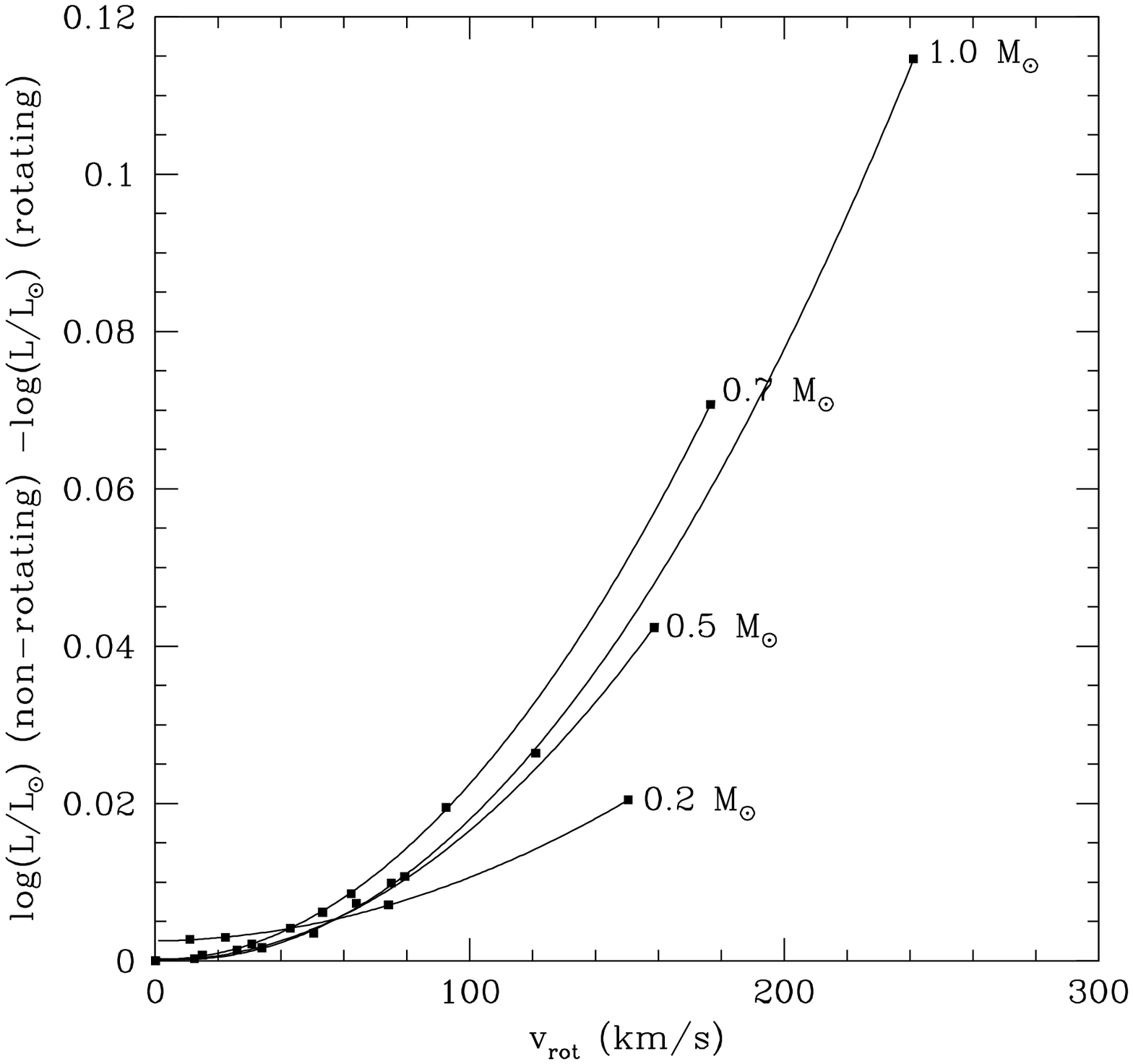]{Difference between luminosity of rotating and
non-rotating stars at the zero age main sequence, as a function of
surface rotation velocity. The highest mass stars demonstrate the
largest different in luminosity. However, even for 1 \Msun, the
difference in luminosity is not significant, unlike the difference in
effective temperature caused by rotation.}

\clearpage
\plotone{fig1.eps}

\clearpage
\plotone{tracks.eps}
 
\clearpage
\plotone{solidtracks.eps}

\clearpage
\plotone{kerot.eps}

\clearpage
\plotone{teff.eps}

\clearpage
\plotone{l.eps}

\clearpage

\begin{deluxetable}{cccc}
\tablecaption {Zero Age Main Sequence Information for Rotating Models with an
Initial Period of 8 Days, Differential Rotation and no Angular
Momentum Loss}
\tablehead{
\colhead {Mass (\Msun)} & \colhead {$\log T_{eff}$}  & \colhead
{$\log(L/L_{\odot})$} & \colhead{Age (Myr)} \nl
}
\startdata
0.2 & 3.526 & -2.282 & 890  \nl
0.5 & 3.572 & -1.450 & 280  \nl
0.7 & 3.634 & -0.881 & 230  \nl
1.0 & 3.752 & -0.159 &  27  \nl
\enddata
\end{deluxetable}

\begin{deluxetable}{ccccc}
\tablecaption{Polynomial Coefficients: $T_{eff}^{no rot} -
T_{eff}^{rot} = Av_{rot}^3 + Bv_{rot}^2 + Cv_{rot}+D$ }
\tablehead{
\colhead{Mass (\Msun)} &  \colhead {A} & \colhead{B} & \colhead{C} &
\colhead{D} 
}
\startdata
0.2 &  $-9.43 \times 10^{-7}$ & $1.08 \times 10^{-3}$ & $-7.71 \times 10^{-3}$ & -0.123 \nl 
0.5 & 0.0 & $1.87 \times 10^{-3}$ & $2.98 \times 10^{-2}$ & -0.746  \nl 
0.7 & $-1.49 \times 10^{-5}$ & $8.13 \times 10^{-3}$ & -0.176 & 1.62 \nl 
1.0 & $ 5.56 \times 10^{-6}$ & $4.37 \times 10^{-3}$ & $3.19 \times 10^{-2}$ &  -0.307 \nl
0.7 solid body & $-1.22 \times 10^{-5}$ & $7.43 \times 10^{-3}$ & -0.217 &  2.63 \nl
1.0 solid body & $ 1.02 \times 10^{-5}$ & $1.53 \times 10^{-3}$ &  0.184 & -2.66 \nl
\enddata
\end{deluxetable}

\end{document}